\documentstyle[aps]{revtex}
\begin{document}
\newcommand {\be}{\begin{equation}}
\newcommand {\ee}{\end{equation}}
\newcommand {\bea}{\begin{array}}
\newcommand {\cl}{\centerline}
\newcommand {\eea}{\end{array}}

\baselineskip 14pt
\title{String Theory, Matrix Model, and  Noncommutative Geometry}
\author{F. Ardalan
\footnote{ E-mail:  ardalan@theory.ipm.ac.ir }} 
\address{Institute for studies in theoretical Physics and Mathematics 
IPM,\\P.O.Box 19395-5531, Tehran, Iran}
\author{}
\address{Department of Physics Sharif University of Technology,\\ 
P.O.Box 11365-9161, Tehran, Iran}
\maketitle

\begin{abstract}
Compactification of Matrix Model on a Noncommutative torus is obtained from
strings ending on D-branes with background B field. The BPS spectrum of the
system and a novel $SL(2,Z)$ symmetry are discussed.
\
\end{abstract}

Noncommutativity of space-time coordinates emerged in string theory recently 
in the context of coincident D-branes [1]; in fact the embedding 
coordinates of D-branes turned out to be noncommutative.
These noncommutative coordinates in the case of 0-branes are elevated to the 
dynamical variables of Matrix Theory, which is conjectured to describe the 
strong coupling limit of string theory, or M-theory, in the infinite momentum 
frame [2].
 
Another kind of noncommutativity in spce coordinates has been recently 
observed in Matrix Theory which is superficially different from the above kind.
It comes from the application of the non-commutative geometry (NCG) techniques
pioneered by A. Connes to the Matrix Theory compactifications [3].  

As a formulation of M-theory, Matrix Theory must describe string theory 
when compactified on a circle; further compactifications being neccessary to 
accomodate low energy physics. A class of toroidal compactifications have 
been known , which relies on a certain commutative subalgebra of matrices 
[4,5]. The subalgebra being an equivalent description of the manifold 
of torus on which compactification is performed.

It was observed by  Connes, Douglas and Schwarz (CDS) that a nonabelian 
generalization of this algebraic description of the manifold of 
compactification, in the spirit of NCG, it is possible to arrive at a 
different compactification of Matrix-model, with the subsequent novel 
physical result of appearance of a constant background of the 3-form field
in the 11 dimensional supergravity limit.

It was immediately observed by Douglas and Hull [6] that a consequent deformed
SYM theory and, therefore indirectly, the noncommutative torus (NCT) 
compactification is a natural consequence of certain D-brane configurations 
in string theory. The subject has been pursued in recent works [7,8,9,10,11].

Thus there is a close connection between constant background Kalb-Ramond field
B and the nonabelian torus  compactification of the Matrix Theory. But, it is 
not obvious how a background B field can make the coordinates noncommutative
and how this noncommutativity differs from that of the coincident D-branes.

We will show explicitly how the CDS noncommutativity arises from D-branes in 
the presence of B field backgraound and compare it with the noncommutativity 
due to coincident D-branes [12,13,21]. This noncommutativity persists in higher 
tori.
The dynamical variables of Marix Theory are $N\times N$ matrices which are 
function of time, with $N$ going to infinity and with the supersymmetric 
action,

\be
I={1 \over 2g\sqrt{\alpha'}}\int d\tau \;\;\;Tr \; \biggl\{ \dot{X_a}\dot{X_a}
+{1 \over {(2\pi\alpha')}^2}\sum_{a<b}[X^a,X^b]^2 \\
+{i \over {2\pi\alpha'}} \Psi^T\dot{\Psi}- {1 \over {(2\pi\alpha')}^2} \Psi^T
\Gamma_a[X^a,\Psi]\biggr\}.
\ee
$X^a, a=1,...,9$ are bosonic hermitian matrices and $\Psi$ are 16 component
spinors. $\Gamma^a$ are $SO(9)$ Dirac matrices. Classical time independent 
solutions have commuting $X^a$, therefore simultaneously diagonalizable, 
corresponding to the  classical coordinates of $N$ 0-branes. In general 
off-diagonal elements of $X^a$ correspond to substringy noncommutative 
structure of M-theory.

Compactification of coordinates $X_i$ of Matrix Theory on a space-like torus 
of radii $R_i$ has been shown [5] to require existence of the matrices $U_i$ 
with the property 

\be
\bea{cc}
U_iX_iU_i^{-1}=X_i+R_i \\
UX^a U^{-1}=X^a  \;\;\;\;\;\; a\neq 1,2\\
U\Psi U^{-1}=\Psi
\eea
\ee
Consistency between these equations requires:
\be
U_iU_j=e^{i\theta_ij}U_jU_i,
\ee
for real numbers $\theta_ij$; where for the usual commutative torus, 
$\theta_ij=0$. In fact {\it rational} $\theta_ij$ will 
also give a commutative torus. It is easily seen that for $\theta=0$,
\be \bea{cc}
X_i=i{\partial \over \partial \sigma_i}+ A_i \;\;\;\; , i=1,...,p \\
U_i=e^{i\sigma_i R_i}.
\eea \ee
is a solution of eq. (2) and (3) and its insertion in the action results 
in the p+1 dimensional SYM on the dual torus. Here $\sigma_i$ parameterize the 
dual torus. 

In the case of two torus, Connes, Douglas and Schwarz [3] observed that in 
Eq. (3), $\theta$ can be taken different from zero and it corresponds to 
compactification on a noncommutative torus (NCT) and the resulting gauge 
theory is the SYM with the commutator of the gauge fields replaced by the 
Moyal bracket. In NCG the $c^*$ algebra of functions over the manifold, is 
generalized to a noncommutative $c^*$ algebra [14]. Thus, the algebra 
generated by the commuting matrices $U_1$ and $U_2$ in the case of usual 
$T^2$, is generalized to the algebra generated by $U_1$ and $U_2$ 
satisfying the relation (3), which now defines a "noncommutative" torus, 
$T^2_{\theta}$. 
The solutions of (3) are then,
\be
X_i=iR_i\partial_i +A_i,
\ee
where $A_i$ now are functions of $\tilde{U_i}$, with $\tilde{U_i}$ satisfying 
\be
\bea{cc}
\tilde{U_1}\tilde{U_2}=\;\;\; 
e^{-i\theta}\tilde{U_2}\tilde{U_1}$, $\;\;\;\;U_i\tilde{U_j}=\tilde{U_j}U_i \\  

 [\partial_i, \tilde{U_j}]=i\delta_{ij} \tilde{U_j} ;\;\; i,j=1,2.
\eea
\ee
Substituting them in the action, we get the SYM theory on the NCT dual to the  
original one, with the essential modification being, the replacement of 
commutators of gauge fields by the Moyal bracket,
\be\bea{cc}
\bigl\{ A,B\bigr\}=A*B-B*A ,\\
A*B(\sigma)=e^{-i\theta(\partial'_1\partial''_2-\partial'_2\partial''_1)} A(\sigma')B(\sigma'')|_
{\sigma'=\sigma''=\sigma}.
\eea \ee
with $\sigma=(\sigma_1,\sigma_2)$.

The BPS spectrum of the compactified Matrix Theory on the noncommutative 
torus has been calculated[3,15], and is, 
\be \bea{cc}
E={R \over {n-m\theta}}\biggl\{ {1 \over 2}\bigl( {n_i-m_i\theta\over R_i}\bigr)^2
+{V^2 \over 2}\bigl[ m+(n-m\theta)\gamma\bigr]^2 \\
+2\pi \sqrt{(R_1w_1)^2+(R_2w_2)^2} \biggr\}. 
\eea\ee
where $V=(2\pi)^2R_1R_2$ and ${n_i \over R_i}$ are KK momenta conjugate to 
$X_i$; $m_i=\epsilon_{ij}m_{j-}$, with $m_{i-}$ winding number of the longitudinal 
membrane along $X_i$ and $X_-$ direction; $R$ the compactification radius 
along the $X_-$ direction and $w_i$ are the momenta of BPS states due to the 
transverse coordinates and are constrained by:
\be
w_i=\epsilon_{ij}(nm_j-mn_j).
\ee
$n$ is the dimension of matrices, $m$ is the winding number of  
the membrane around torus and $\theta$ is the deformation parameter of the 
torus. The mass spectrum (8) is invariant under an $SL(2,Z)_N$ generated by
\be\bea{cc}
\theta \rightarrow {-1 \over \theta} \\
m \rightarrow n \;\;\;\; , \;\;\;\; n \rightarrow -m \\
m_i \rightarrow n_i \;\;\; , \;\;\;\; n_i \rightarrow -m_i \\
\gamma \rightarrow -\theta(\theta\gamma+1) \\
R_i \rightarrow \theta^{-2/3} R_i \;\;\; , \;\;\;\;\; R \rightarrow 
\theta^{-1/3} R
\eea\ee
and
\be\bea{cc}
\theta \rightarrow  \theta+1 \\
n \rightarrow n+m \;\;\;\ , \;\;\;\;\; m \rightarrow m \\
n_i \rightarrow n_i+m_{i} \;\;\;\ , \;\;\;\;\; m_i \rightarrow m_i.
\eea\ee
This invariance is to be expected  on the basis of the NCG considerations. It
is the SL(2,Z) invariance of the $c^*$-algebra defining the NCT [3].

We will now see how the above noncommutativit appears in string theory 
in the presence of D-branes in the $B_{\mu\nu}$ background. The dynamics of 
strings ending on a p-brane in the background of the antisymmetric field, 
$B_{\mu\nu}$ is [15],

\be\bea{cc}
S= {1 \over 4\pi\alpha'} \int_{\Sigma} d^2\sigma \bigl[ \eta_{\mu\nu}
\partial_aX^{\mu}\partial_bX^{\nu}g^{ab}+ \epsilon^{ab} B_{\mu\nu}\partial_a
X^{\mu}\partial_bX^{\nu}+ 
{1 \over 2\pi\alpha'}\oint_{\partial \Sigma} d \tau A_i \partial_{\tau}\zeta^i, 
\eea\ee
where $A_i,\ i=0,1,p$ is the $U(1)$ gauge field living on the D-brane and
$\zeta^i$ its internal coordinates. 
The action is invariant under the combined gauge transformation [1]
\be \bea{cc}
B_{\mu\nu}\rightarrow B_{\mu\nu}+\partial_{\mu}\Lambda_{\nu}-\partial_{\nu}\Lambda_{\mu} \\
A_{\mu} \rightarrow A_{\mu}-\Lambda_{\mu}.
\eea\ee
The gauge invariant field strength is then 
\be
{\cal F}_{\mu\nu}=B_{\mu\nu}-F_{\mu\nu} \;\;\;\;\; ,\;\; 
F_{\mu\nu}=\partial_{[\mu}A_{\nu]}.
\ee
which leads to the following mixed boundary conditions, 

\be
\left\{  \begin{array}{cc}
\partial_{\sigma}X_0=0 \\
\partial_{\sigma}X_i+{\cal F}_{ij} \partial_{\tau}X_j=0 \\
\partial_{\sigma}X_i- {\cal F}_{ij} \partial_{\tau}X_j=0  \;\;\; \\
\partial_{\tau} X_a=0 \;\;\;\ ,\;\;\; a=p,...,9.
\end{array}\right.
\ee
Canonical commutation relations of $X_i$ and their conjugate momenta
$P_i$, $i=1,...,p$:
\be 
P_i=\partial_{\tau}X_i-{\cal F}_{ij}\partial_{\sigma}X_j \;\;\; ,\;\;\; 
\ee
\be
[X^{\mu}(\sigma,\tau),P^{\nu}(\sigma',\tau)]=i\eta^{\mu\nu}\delta(\sigma-\sigma').
\ee
Lead to the noncommutative center of mass coordinates:
\be
x^i={1 \over \pi}\int X^i(\sigma,\tau)\ d\sigma,
\ee
\be
[x_i,x_j]=\pi i{\cal F}_{ij}.
\ee

This noncommutativity of space coordinates is the reason for the 
noncommutativity which appears in the compactification of Matrix Theory on 
a torus with a constant 3-form field, which can be seen by going to the string
matrix model [16]. To see the connection between the noncommutativity due to 
the boundary conditions on the d-branes and the noncommutativity which appears 
in the transverse coordinates of coincident D-branes, recall that D2-branes 
with a non-zero U(1) gauge field in the background contain a distribution of 
0-branes proportional to ${\cal F}$ [12,13,17,18,19]; thus the noncommutativity 
of the coordinates. 

It is interesting that the mechanism which produces the original 
noncommutativity in the description of D-branes, and leads through a set of 
arguments to the particular form of the commutation relation in (19), is 
simply derived from the string action (12) in the presence of the ${\cal F}$ and mixed boundary conditions with $B$ field background.

We compactify the $X^i$ direction and wrap the 2-brane around the 
2-torus and use the center of mass coordinates $x^i$ and their conjugate 
momenta to construct the generators of the $c^*$ algebra of the noncommutative
torus; proving that the compactification, in the presence of $U(1)$ field 
strength, for D-membrane requires a NCT;
\be
\bea{cc}
U_1x^1U_1^{-1}=x^1+ R_1 \\
U_2x^2U_2^{-1}=x^2+ R_2 \\
U_ix^j U_i^{-1}=x^j  \;\;\;\;\;\; i\neq j=1,2
\eea
\ee
A solution to these equations is:
\be\bea{cc}
U_1=exp\{ -i R_1\bigl[a(p_1- {x^2 \over \pi{\cal F}})- {x^2 \over 
\pi{\cal F}}\bigr]\}\\
U_2=exp\{-i R_2\bigl[a(p_2+{x^1 \over {\cal F}})+ {x^1 \over 
{\cal F}}\bigr]\},
\eea\ee
with $a^2=1+{\pi^2{\cal F}^2\over R_1R_2}$. The above relations leads to
\be
U_1U_2 =e^{i\pi {\cal F}} U_2U_1.
\ee
This result reproduces the Matrix Theory compactification on the NCT
formulated by CDS, described previously. It was argued there that, the 
noncommutativity of the torus is related to the non-vanishing of 3-form of 
M-theory, which in the string theory reduces to the antisymmetric NSNS 
2-form field, $B_{\mu\nu}$. In our case noncommutativity of the torus on
which the D-membrane of string theory is compactified, is a direct result of  
the non-vanishing $B$ field. In fact using the Matrix model formulation
of string theory [16]. it is straightforward to obtain CDS results.

The noncommutativity of the $c^*$ algebra (20) and (3) of the NCT is
similar to, but distinct from, the noncommutativity of the coordinates as in 
(19) and as it appears in Matrix Theory and bound states of D-branes. 
The similarities are obvious, but the differences are subtle. In fact it is 
possible to see that when ${\cal F}$ is quantized to a rational number, by an
SL(2,Z) transformation, we can make the $U_1$ and $U_2$ commute, i.e. 
we can make the torus {\it commutative}, while the coordinates are 
{\it noncommutative}. 
Thus for {\it irrational} parameter $\theta$, we are dealing with a new 
form of noncommutativity not encountered in ordinary Matrix theory or
in the context of D-brane bound state.

We will now find the BPS spectrum of a system of (D2-D0)-brane bound state.
It is convenient to consider the T-dual version of the mixed brane, in which 
we only need to deal with commutative coordinates and commutative torus, 
and are able to calculate the related spectrum just by theusual string 
theory methods.

Applying T-duality in an arbitrary direction, say $X^2$,
\be
\left\{  \begin{array}{cc}
\partial_{\sigma}X^0=0 \\
\partial_{\sigma}(X^{1}+{\cal F} X^2)=0 \\
\partial_{\tau}(X^{2}- {\cal F} X^{1})=0  \;\;\; \\
\partial_{\tau} X^a=0 \;\;\;\ ,\;\;\; a=3,...,9,
\end{array}\right.
\ee
describing a tilted D-string which makes an angle $\phi$ with the
duality direction, $X^2$: 
$$
\cot \phi={\cal F}.
$$
Thus we consider a D-string winding around a cycle of a torus defined by:
\be
\tau={R_2 \over R_1}e^{i\alpha}=\tau_1 +i \tau_2 \;\;\;\; , \;\;\;\;\
\rho=iR_1R_2\sin \alpha+b=i\rho_2+b,
\ee
where $b=BR_1R_2\sin\alpha$ is the flux of the $B$ field on the torus.
The D-string is located at an angle $\phi$ with the $R_1$ direction such that it 
winds $n$ times around $R_1$ and $m$ times around $R_2$. Hence
\be
\cot \phi={n \over m \tau_2}+\cot \alpha.
\ee

The BPS spectrum of this tilted D-string system gets contributions from both
the open strings attached to the D-string and the D-string itself. 
The open strings have mode expansions [16]: 

\be\left\{ \bea{cc}
X^i=x_0^i + p^i \tau+ L^i \sigma + Oscil. \;\;\; , i=1,2  \\
X^0=x^0_0+p^0\tau+ Oscil. \\
X^a=x^a_0+ Oscil. \;\;\; , \;\;\; a=3,...,9 
\eea\right.
\ee
where $p^i$ and $L^i$, in usual complex notation, are:

\be 
p=\ r_1 {{n+m\tau} \over {|n+m\tau|^2}} \sqrt{{\tau_2\over \rho_2}}
\;\;\;\;\;\ ; r_1\in Z.
\ee
\be
L=\ q_1 {\rho{(n+m\tau)} \over {|n+m\tau|^2}} \sqrt{{\tau_2\over \rho_2}}
\;\;\;\;\;\ ; q_1\in Z.        
\ee

Mass of the open string is then,
\be 
M^2=|p+L|^2+{\cal N}= {{\tau_2}\over {|n+m\tau|^2}}{{|r_1+q_1\rho|^2} \over 
\rho_2}+{\cal N},
\ee
where ${\cal N}$ is the contribution of the oscillatory modes.
This mass is invariant under both $SL(2,Z)$'s of the torus
acting on $\rho$ and $\tau$. Applying T-duality in $R_1$ direction,
$$
R_1 \rightarrow {1 \over R_1}\;\;\;\;\; {\rm or\ \ equivalently} 
\;\;\;\tau\leftrightarrow \rho,
$$
we obtain the spectrum of the open string compactified on NCT,

\be 
M^2={{\rho_2}\over {|n+m\rho|^2}}{{|r_1+q_1\tau|^2} \over \tau_2} 
+{\cal N},
\ee

Next we consider the D-string contribution. For this purpose we use the DBI 
action [9,20],
\be
S_{D-string}={-1\over g_s}\int d^2\sigma \sqrt{det(\eta_{ab}+{\cal F}_{ab})}.
\ee
with
\be
\eta_{ab}=\left (\bea{cc}1-v^2 \;\;\;\;\ 0 \\ 0 \;\;\;\;\;\;\;\;\;\;  1  
\eea\right),
\ee
\be
{\cal F}_{ab}=\left (\bea{cc} 0 \;\;\;\;\ Bv+F \\ Bv+F \;\;\;\;\;\  0
\eea\right).
\ee
This action leads to the masss pectrum
\be
\alpha' M^2={|n+m\tau|^2 \rho_2 \over \alpha' g_s^2 \tau_2}+\alpha' {|r_2+\rho q_2|^2 \over
\rho_2 \tau_2}.
\ee
Applying T-duality, we find,
\be
\alpha' M^2_{membrane}={|n+m\rho|^2 \tau_2 \over \alpha'  {g'_s}^2 \rho_2} 
+\alpha' {|r_2+\tau q_2|^2 \over\rho_2\tau_2}.
\ee
The $SL(2,Z)_N$ invariance, acting on $\rho$, is manifestly seen from the 
above equation. The open strings and the D-string form a {\it marginal} bound 
state, and the full BPS spectrum is the addition of the separate contributins,
\be
{\cal M}= M_{membrane} + M_{open\  st.}.
\ee

\be
{\cal M}=\sqrt{{\tau_2 \over \rho_2}}{|n+m\rho| \over g'_s}
(1+{g'_s}^2{|r_2+q_2\tau|^2 \over \tau_2}{\rho_2 \over |n+m\rho|^2})^{1/2}+ 
{|r_1+q_1\tau|\over |n+m\rho|}\sqrt{{\rho_2 \over \tau_2}}.
\ee
The above spectrum is manifestly $Sl(2,Z)_N$ invariant,in the notation of CDF.
In the zero volume and $g_s \rightarrow 0$ limits,

\be
l_s{\cal M}={|n-m\theta| \over g_s} +{1 \over 2g_s}{m^2V^2 \over |n-m\theta|}
+{g_s \over 2|n-m\theta|}{|r_2+q_2\tau|^2 \over \tau_2}+
{|r_1+q_1\tau|\over |n-m\theta|}\sqrt{{V \over \tau_2}}.
\ee

The $SL(2,Z)_N$ symmetry generators are 
$$
\rho \rightarrow \rho+1 \;\;\;\;\; ,\;\;\;\;\;\; \rho\rightarrow {-1\over \rho}
$$
which in the zero volume limit ($\rho_2=0$) become
\be
\theta\rightarrow \theta+1 \;\;\;\ , \;\;\; \theta\rightarrow {-1\over \theta}
\ee

Invariance of the mass spectrum, under $\theta\rightarrow 
{-1\over \theta}$, implies that
\be
g_s\rightarrow g'_s=g_s {\theta}^{-1} 
\ee
Moreover the imaginary part of $\rho\rightarrow {-1\over \rho}$, tells us that 
the volume of the torus in the zero volume limit, in the string theory units, 
transforms as:
\be
V\rightarrow V'=V \theta^{-2}
\ee
Putting these relations together, and remembering the relation of 
10 dimensional units and 11 dinemsional parameters, $l_p^3=l_s^3 g_s$ 
and $l_sg_s=R$, and assuming $l_p$  invariance under $\theta$  
transformations, we obtain: 
\be\bea{cc}
R \rightarrow R'=R {\theta}^{-2/3}  \\
R_i\rightarrow R'_i=R_i {\theta}^{-2/3} \\
l_s\rightarrow l'_s=l_s {\theta}^{-1/3}. 
\eea\ee
The above relations indicate an M-theoretic origin for the $SL(2,Z)_N$. 
This is the effect of considering the whole DBI action and not, 
only its second order terms[20]


\begin{references}
\bibitem{}  E. Witten,  Nucl. Phys. {\bf B460}, 335 (1996).
\bibitem{}  T. Banks, W. Fischler, S.H. Shenker, and L. Susskind, 
Phys. Rev. {\bf D55}, 5112 (1997).
\bibitem{}
A. Connes, M.R. Douglas, A. Schwarz,
"Noncommutative Geometry and Matrix Theory: Compactification on Tori",
JHEP {\bf 02}, 003 (1998), hep-th/9711162.
\bibitem{} W. Taylor IV, hep-th/9801182 and references therein.
\bibitem{} W. Taylor IV,  Phys. Lett. {\bf B394}, 283 (1997), 
hep-th/9611042.
\bibitem{}
O. Ganor, S. Ramgoolam and W. Taylor IV,  Nucl. Phys. {\bf B492}, 191
(1997), hep-th/9611202. 
\bibitem{} M. R. Douglas, C. Hull, "D-branes and Noncommutative Torus",
JHEP {\bf 02} 008 (1998), hep-th/9711165.
\bibitem{}
M. Berkooz, "Non-local Field Theories and the Noncommutative Torus",
hep-th/9802069.
\bibitem{} P.-M. Ho, Y.-S. Wu,
"Noncommutative Gauge Theories in Matrix Theory", hep-th/9801147; C. S. Chu
and P. M. Ho, ``Noncommutative Open Strings and D-branes'', hep-th/9812219;
P. M. Ho, Phys. Lett. {\bf B434}, 41 (1998).
\bibitem{} M. Li,
"Comments on Supersymmetric Yang-Mills Theory on a Noncommutative Torus",
hep-th/9802052.
\bibitem{}
Y.-K. E. Cheung, M. Krogh,
"Noncommutative Geometry From $0$-Branes in a Background $B$ Field",
Nucl. Phys. {\bf B528}, 185 (1998), hep-th/9803031.
\bibitem{}  
T. Kawano and K. Okuyama,  "Matrix Theory on Noncommutative Torus", 
hep-th/9803044. 
\bibitem{} F. Ardalan, H. Arfaei, M. M. Sheikh-Jabbari,
"Mixed Branes and Matrix Theory on Noncommutative Torus", Proceeding of 
PASCOS 98, hep-th/9803067.
\bibitem{} F. Ardalan, H. Arfaei, M. M. Sheikh-Jabbari, JHEP {\bf 02}, 016 
(1999), hep-th/9810072.
\bibitem{}	
A.Connes, Noncommutative Geometry, Academic Press, 1994.
\bibitem{}  R. G. Leigh, Mod. Phys. Lett. {\bf A4}, 28, 2767 (1989). 
\bibitem{} R. Dijkgraaf, E. Verlinde, H. Verlinde, Nucl. Phys. {\bf
B500}, 43 (1997).
\bibitem{}  M.M. Sheikh-Jabbari, "More on Mixed boundary Conditions 
and D-brane Bound States", Phys. Lett. {\bf B425}, 48 (1998),
hep-th/9712199.
\bibitem{}  H. Arfaei and M.M. Sheikh-Jabbari, 
Nucl. Phys. {\bf B526}, 278 (1998), hep-th/9709054.

\bibitem{}  H. Arfaei and M.M. Sheikh-Jabbari,  Phys. Lett. 
{\bf B394} 288 (1997).
\bibitem{} C. Hofman, and E. Verlinde, JHEP {\bf 12}, 010 (1998). 
\bibitem{} M. M. Sheikh-Jabbari, ``Renormalizability of the Supersymmetric
Yang-Mills Theories on the Noncommutative Torus'', hep-th/9903107.
\end{references}
\end{document}